\documentclass[useAMS,usenatbib]{mn2e}
\usepackage{graphicx}

\title[Millimetric atmospheric emission at Dome Concordia]{Intensity and polarization of the atmospheric emission at millimetric wavelengths at Dome Concordia}
\author[E.S. Battistelli et al.]{ E.S.
Battistelli,$^{1}$\thanks{E-mail: elia.battistelli@roma1.infn.it}
G. Amico,$^{1}$ A. Ba\`{u},$^{2}$ L. Berg\'{e},$^{3}$ \'{E.}
Br\'{e}elle,$^{4}$ R. Charlassier,$^{4}$ \newauthor S.
Collin,$^{3}$ A. Cruciani,$^{1}$ P. de Bernardis,$^{1}$ C.
Dufour,$^{4}$ L. Dumoulin,$^{3}$ \newauthor M. Gervasi,$^{2}$ M.
Giard,$^{5}$ C. Giordano,$^{1,6}$ Y. Giraud-H\'{e}raud,$^{4}$ L.
Guglielmi,$^{4}$ \newauthor J.-C. Hamilton,$^{4}$ J.
Land\'{e},$^{5}$ B. Maffei,$^{7}$ M. Maiello,$^{1,8}$ S.
Marnieros,$^{3}$ S. Masi,$^{1}$ \newauthor A. Passerini,$^{2}$ F.
Piacentini,$^{1}$ M. Piat,$^{4}$ L. Piccirillo,$^{7}$ G.
Pisano,$^{7}$ G. Polenta,$^{1,9,10}$ \newauthor C. Rosset,$^{4}$
M. Salatino,$^{1}$ A. Schillaci,$^{1}$ R. Sordini,$^{1,11}$ S.
Spinelli,$^{2,12}$ A. Tartari$^{2,4}$ \newauthor and M.
Zannoni$^{2}$
\\
$^{1}$Dipartimento di Fisica, ``Sapienza'' Universit\`{a} di Roma,
Piazzale Aldo Moro, 5, 00185, Rome, Italy
\\
$^{2}$Dipartimento di Fisica ``G.Occhialini'', Universit\`{a}
degli Studi di Milano-Bicocca, Piazza della Scienza, 3, 20126,
Milan, Italy
\\
$^{3}$Centre de Spectroscopie Nucl\'{e}aire et de Spectroscopie de
Masse, UMR8609 IN2P3-CNRS, Universit\'{e} Paris Sud, b\^{a}t 108,
91405, \\ Orsay Campus, France
\\
$^{4}$APC, Universit\'{e} Paris, Diderot-Paris 7, CNRS/IN2P3, CEA,
Observatoire de Paris, 10, rue A. Domon \& L. Duquet, Paris,
France
\\
$^{5}$Centre d'\'{E}tude Spatiale des Rayonnements,
CNRS/Universit\'{e} de Toulouse, 9 Avenue du colonel Roche, BP
44346, 31028, Toulouse \\ Cedex 04, France
\\
$^{6}$Fondazione Bruno Kessler, Via S.Croce 77, 38122, Trento,
Italy
\\
$^{7}$JBCA School of Physics and Astronomy, The University of
Manchester, Alan Turing Building, Oxford Road, Manchester M13 9PL,
UK
\\
$^{8}$Universit\`{a} degli Studi di Siena, Via Banchi di Sotto 55,
53100, Siena, Italy
\\
$^{9}$ASI Science Data Center, c/o ESRIN, via G. Galilei, 00044,
Frascati, Italy
\\
$^{10}$INAF - Osservatorio Astronomico di Roma, via di Frascati
33, 00040 Monte Porzio Catone, Italy
\\
$^{11}$Dipartimento di Scienze Applicate, Universit\`{a} degli
Studi di Napoli ``Parthenope'', Centro Direzionale di Napoli,
Isola C4, 80143, \\ Naples, Italy
\\
$^{12}$Media Lario Technologies S.r.l., Localit\`{a}  Pascolo,
23842, Bosisio Parini (LC), ITALY}

\begin{document}

\date{Received ...; accepted ...}

\pagerange{\pageref{firstpage}--\pageref{lastpage}} \pubyear{2012}

\maketitle

\label{firstpage}

\begin{abstract}
Atmospheric emission is a dominant source of disturbance in
ground-based astronomy at millimetric wavelengths. The Antarctic
plateau is recognized to be an ideal site for millimetric and
sub-millimetric observations, and the French/Italian base of Dome
Concordia is among the best sites on Earth for these observations.
In this paper we present measurements at Dome Concordia of the
atmospheric emission in intensity and polarization at $2\ mm$
wavelength, one of the best observational frequencies for Cosmic
Microwave Background (CMB) observations when considering cosmic
signal intensity, atmospheric transmission, detectors sensitivity,
and foreground removal. Using the BRAIN-pathfinder experiment, we
have performed measurements of the atmospheric emission at $150\
GHz$. Careful characterization of the air-mass synchronous
emission has been performed, acquiring more that 380 elevation
scans (i.e. ``skydip'') during the third BRAIN-pathfinder summer
campaign in December 2009/January 2010. The extremely high
transparency of the Antarctic atmosphere over Dome Concordia is
proven by the very low measured optical depth: $<\tau_{I}>=0.050
\pm 0.003 \pm 0.011$ where the first error is statistical and the
second is systematic error. Mid term stability, over the summer
campaign, of the atmosphere emission has also been studied.
Adapting the radiative transfer atmosphere emission model {\em am}
to the particular conditions found at Dome Concordia, we also
infer the level of the precipitable water vapor (PWV) content of
the atmosphere, notoriously the main source of disturbance in
millimetric astronomy ($<PWV>=0.77 \pm 0.06 \pm ^{0.15}_{0.12}\
mm$). Upper limits on the air-mass correlated polarized signal are
also placed for the first time. The degree of circular
polarization of atmospheric emission is found to be lower than
0.2\% (95\%CL), while the degree of linear polarization is found
to be lower than 0.1\% (95\%CL). These limits include
signal-correlated instrumental spurious polarization.
\end{abstract}


\section{Introduction}

Fast growing fields in millimetric astronomy are the study of the
Cosmic Microwave Background (CMB) polarization and the measurement
of polarized emission from interstellar dust. In particular, a
curl component (B-modes) of the CMB polarization from the
predicted inflationary expansion of the universe earlier in time
may be present. The theorized signal depends on the energy of the
inflationary field, as measured by the tensor-to-scalar ratio r,
and is so low that exquisite sensitivity and control of systematic
effects are necessary to attempt these kind of observations.

Atmospheric emission is one of the dominant sources of disturbance
for ground-based CMB experiments and for millimetric and
sub-millimetric astronomy in general. In addition to continuum
emission at the frequencies of our interest, there are also the
roto-vibrational emission lines of $O_{2}$ (at around $60\ GHz$
and $119\ GHz$) and $H_{2}O$ (at $22\ GHz$ and $183\ GHz$). Dry
and high altitude observation sites are chosen to mitigate the
problem. The French/Italian scientific base of Dome Concordia on
the Antarctic plateau ($75^{\circ}06'$ South, $123^{\circ}24'$
East, at $3233\ m$ aside see level, http://www.concordiabase.eu/)
is one of the best observational sites on Earth for millimetric
observations. Site testing at Dome C has proven its observational
quality at different wavelengths (see e.g.
\cite{tremblin2011,gredel2010,lawrence2004,aristidi2009,calisse2004})
although only preliminary measurements were performed at
millimeter wavelengths \cite{valenziano1999}. $150\ GHz$ is among
the best observational frequencies for ground based CMB
experiments in terms of cosmic signal intensity, atmospheric
transmission, detectors sensitivity and foreground removal.

The BRAIN-pathfinder experiment \cite{masi2005,polenta2007} has
undergone its third Antarctic campaign from the French/Italian
scientific base of Dome Concordia. The first two campaigns were
dedicated to instrument fielding, while the 2009-2010 austral
summer campaign was dedicated to continuous observations of the
atmospheric emission and site testing. This paper is structured as
follows: in $\S$2 we introduce the BRAIN-pathfinder instrument, in
$\S$3 we describe the observations; in $\S$4 we present the data
and the analysis and in $\S$5 we give the results in terms of
intensity and polarization.


\section{BRAIN-pathfinder: the instrument}

The BRAIN-pathfinder was designed as a prototype instrument for a
challenging project of bolometric interferometer. The BRAIN
collaboration has been combined with the Millimeter-wave
Bolometric Interferometer (MBI) \cite{tucker2008} collaboration to
form the QUBIC collaboration. The BRAIN-pathfinder was devoted to
site and logistics testing for the QUBIC experiment \cite{qubic},
that we aim to install at Dome Concordia in 2013. The
BRAIN-pathfinder instrument is described in detail elsewhere
\cite{masi2005}; \cite{polenta2007}; \cite{braincrio}.
Nevertheless, we here report a brief description of the
instrumental setup for completeness.

The BRAIN-pathfinder comprises of a two-channel bolometric
receiver coupled to two off-axis ($40$ and $60\ cm$ diameter)
parabolic mirrors, tiltable around the optical axis. The whole
instrument is mounted on an azimuth plane, making the instrument
an Alt-Az double telescope. The first in its kind, our bolometric
receiver is cooled by a dry-cryostat with a Sumitomo
\footnote{http://www.shicryogenics.com/} Pulse Tube cryocooler,
allowing us to keep an intermediate stage at $30\ K$ and a main
plate at $3\ K$. Quasi-optical filters, JFET boards and shields
are kept at $30\ K$ with JFET amplifiers attached to their PCB
through weak thermal connections in their fiberglass supports.
Further filters, radiation collecting horns and further shields
are kept at $3\ K$ with an $^{3}He-^{4}He$ refrigerator
\footnote{http://www.chasecryogenics.com/} that keeps the
bolometers at 310mK during observations.

One of the two channels (channel 1) measures the anisotropy of the
emission of the sky, while the second one (channel 2) sees the sky
through an ambient-temperature, rotating sapphire Quarter Wave
Plate (QWP), and a steady wire grid polarizer. The presence of the
rotating QWP makes this second detector inherently sensitive to
linear and circular polarization. In an ideal case, the power
hitting channel 2 is thus \cite{polenta2007}:
\begin{equation}\label{eq:1}
\centering W=\frac{1}{2}[I_{in}+Q_{in}\frac{1+cos(4 \omega
t)}{2}+U_{in}\frac{sin(4 \omega t)}{2}+V_{in} sin(2 \omega t)]
\end{equation}
where $I_{in}$, $Q_{in}$, $U_{in}$, and $V_{in}$ and the Stokes
parameters of the incoming radiation, $\omega$ is the mechanical
angular speed of the QWP. From equation \ref{eq:1} it is clear
that an incoming polarized radiation is modulated by the rotating
QWP at a frequencies twice or four times the mechanical frequency,
depending whether the polarization is circular or linear.

We use the control and read-out electronics, as well as the
control software, originally developed for the Planck High
frequency Instrument \cite{lamarre2010} and for its ground based
calibrations. Its angular resolution on the sky is $1^{\circ}$. A
double back-to-back horn and the quasi-optical filters set the
average observational frequency at $150\ GHz$, with $36\ GHz$ FWHM
bandwidth. Detailed measurements of the bandpass have been
performed combining data obtained with a high throughput Fourier
Transform Spectrometer \cite{schillophd} and a Vector Network
Analyzer \footnote{http://www.home.agilent.com/agilent/home.jspx}
in order to characterize the transmission curve especially on the
low frequency end of the band-pass, where the molecular Oxygen
line emission becomes brighter.

\section{Observations}

Observations were taken during the 2009-2010 austral summer
Antarctic campaign. As opposed to a CMB experiment, for which one
should choose an observational strategy aiming at minimizing the
atmospheric effects (see eg. \cite{chiang2010},
\cite{castro2009}), we have chosen an observational strategy able
to highlight the atmospheric contribution. Here we present a full
characterization of the air-mass dependence of the atmospheric
intensity and polarized emission, obtained by performing elevation
scans (i.e. skydips) by leaving the azimuth constant and scanning
the elevation from the zenith to $35$ degrees above the horizon.

Elevation scans were done in a so called ``fast scan'' mode by
acquiring the sky signal while the telescope continuously samples
different elevation angles. The scan speed has been chosen as high
as possible, in order to mitigate the $1/f$ noise present in the
data arising both from detector instability and (mainly) from the
slow variation of the atmospheric emission. At the same time we
set the scan speed to be able to acquire multiple QWP rotations in
one single telescope beam. We tested different QWP and scan speeds
and, trading off instrumental constrains and observational needs.
We set the QWP rotational frequency at $1.56\ Hz$ (corresponding
to $3.13\ Hz$ and $6.26\ Hz$ respectively for circular and linear
polarization modulation frequencies) for approximately half of the
measurements and at $2.09\ Hz$ (corresponding to $4.17\ Hz$ and
$8.35\ Hz$ respectively for circular and linear polarization
modulation frequencies) for the rest of the time. A scan speed of
$1^{\circ}/s$ thus allows at least three or six polarized
modulation periods per telescope beam. With these settings, a full
scan completes in less than a minute, a short enough time to
mitigate slow signal variation arising from atmospheric emission.

In order to reduce the field of view vignetting, to overcome
detector and read-out non-linearity, and to optimize the detectors
dynamics, we have chosen not to use a warm reference load for
skydip measurements. This forces us to make some assumptions if we
want to do absolute calibration of the data (see section
\ref{sec:reduction}). On the other hand, it allows us to directly
analyze polarization data relatively to the intensity emission,
with no need of any reference signal.

We have collected 383 skydips. About 12\% of the skydips were
discarded because of corrupted data or due to large atmospheric
fluctuations. About 76\% of the analyzed skydips were acquired
changing the elevation from the zenith to $35$ degrees above the
horizon (long skydip) while the remaining 24\% were limited to
$60$ degrees above the horizon (short skydip), in order to keep
the Sun always at an angle larger than $30$ degrees from the field
of view. Most of the reported results have been extracted from the
analysis of the long skydips, although the short ones have been
useful for cross checks.

\section{Data reduction and analysis}\label{sec:reduction}

The skydip technique \cite{dicke1946} is a well investigated
method to study the atmospheric emission (see e.g.
\cite{dragovan1990}, \cite{archibald2002}). During a skydip we
expect the acquired signal to respond to the air-mass as in the
following:

\begin{equation}\label{eq:seclaw}
\centering S=off+C\cdot T_{0}\cdot [1-exp(-x\cdot \tau_{I})]
\end{equation}
where $S$ is the acquired signal in ADC units, $off$ is an
instrumental offset in ADC units, $C$ is the calibration factor in
ADC/K units, $T_{0}$ (in K) is the equivalent temperature of the
atmosphere, $\tau_{I}$ is the sky opacity and $x=sec(z)$ is the
air-mass with the zenithal angle $z=90^{\circ}-elevation$.

One of the features in our data is the presence in the time stream
of a periodic signal at a frequency of around $1\ Hz$ due to the
pulses of the Pulse Tube cryocooler. In fact, the experimental
effort spent to reduce the system vibration has drastically
reduced but not completely eliminated the effect of the pulses on
the high impedence bolometers, that are intrinsically microphonic.
This does not affect the high signal-to-noise skydip measurements
we are analyzing in this paper; nevertheless we have decided to
filter out from our data this well defined imprint, in order to
avoid biases in the skydip fits. We have performed several tests,
both in time and in Fourier space and we finally decided to use a
multiple Notch filter to remove the main pulse frequency and its
harmonics.
\begin{figure}
\begin{center}
\includegraphics[angle=90,width=9cm]{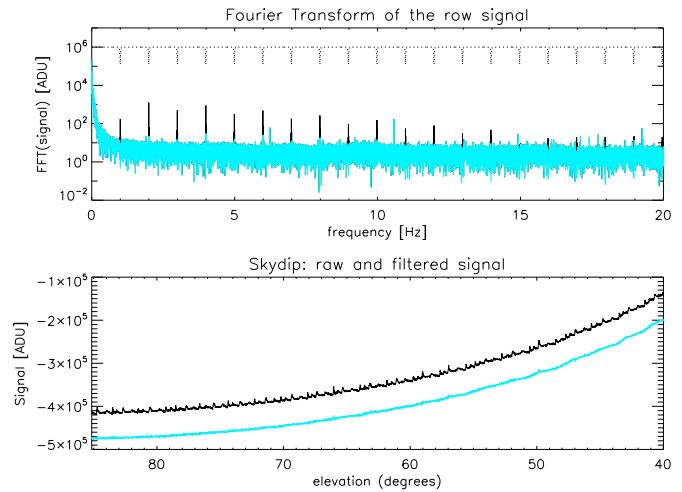}
\caption{Data acquired during one skydip. The plot on the top
shows the Fourier transform of the raw data (black, solid), the
used multiple Notch filter (black, dotted) and the filtered data
(cyan, solid, overplotted on the data). The plot on the bottom
shows a raw skydip (black), and a filtered skydip (cyan) offset by
60000 Analog to Digital Units (ADU).} \label{skydip1}
\end{center}
\end{figure}
In figure \ref{skydip1} we show filtered and unfiltered signals
(and their Fourier transforms), in engineering units, acquired
during one skydip.

While for channel 1 we have only (Notch-) filtered out the Pulse
Tube $1\ Hz$ signal, in order to retrieve signal intensity
information from channel 2 we have to account for the presence of
the rotating QWP in front of it. We have thus applied a low-pass
filter, with cut-off frequency at $1\ Hz$, to remove any higher
frequency signal due to the rotating QWP that could modulate
linear and circular polarized signal. Dedicated band-pass filters
have then been applied to retrieve polarization information (see
section \ref{sec:pol} for details).

We have performed a likelihood analysis and chi square
minimization to optimize secant-law fits on each of the acquired
skydips. We have verified the linearity of the dependence of the
signal as a function of the air-mass. This is verified only in the
case of high transparency of the atmosphere:
\begin{equation}\label{eq:fit}
\centering S = P_{1} + P_{2} \cdot exp(-P_{0} \cdot x)  \simeq
P_{1} + P_{2} - P_{2} \cdot P_{0} \cdot x = A^{I} - B^{I} \cdot x
\end{equation}
where $A^{I} = P_{1} + P_{2}$ and $B^{I} = P_{2} \cdot P_{0}$ are
the fitted parameters.

Comparing equation \ref{eq:seclaw} and \ref{eq:fit} we determine
the sky opacity:

\begin{equation}\label{eq:tau}
\centering \tau_{I} = -B^{I} / (C \cdot T_{0})
\end{equation}
where the calibration factor $C$ has been determined by using
laboratory absolute reference loads cooled at liquid Nitrogen
temperature (i.e. $77\ K$), careful measurements of the bolometer
efficiency, as well as the daily electrical responsivity
measurements performed during observations through the
measurements of the detector $I-V$ curve. The temperature $T_{0}$
was determined by integrating the temperature, pressure and
humidity profiles obtained by daily measurements with atmospheric
radio-sound balloons flying up to an altitude of $24000\ m$,
collecting data with $\Delta h \sim 10\ m$ altitude
sampling\footnote{Data and information were obtained from
IPEV/PNRA Project "Routine Meteorological Observation at Station
Concordia - www.climantartide.it."}. In particular, these data,
combined with continuous ground temperature measurements at the
time of the skydip, and after integration of the atmospheric
emission contribution over the altitude for each of our scans,
allowed us to recover $T_{0}$ using radiative transfer.

Our model for atmosphere emission is made by two different parts:
in order to reconstruct the circularly polarized $O_2$ signals we
use the model described in \cite{spinelli2011}. This model can be
used to estimate also Stokes paramaters I,Q and U. In order to
deal with a complete dry-air model, together with a water vapor
column,  we used the {\em am} model \cite{paine2011}. The {\em am}
model uses updated spectroscopic parameters and is readily
available, very well supported, and documented. We have made
day-by-day estimates of the atmospheric emission using radio
sounds, properly resampled to build {\em am} configuration files.
The resulting brightness temperature is then averaged over the
frequency bandpass of BRAIN-pathfinder channels, after accurate
laboratory bandpass reconstruction, to obtain the power delivered
to our detectors. We find a linear scaling relation between the
precipitable water vapor (PWV), or the integrated optical depth
$\tau$, and the brightness temperature $T_{b}$ in our bandpass,
that is:
\begin{equation}\label{eq:t_b}
\centering T_b=Q + M \cdot PWV,
\end{equation}
where $M_{ch1}= (6.6 \pm 0.3) K/mm$, $M_{ch2}= (5.4 \pm 0.3)
K/mm$, $Q = (4.2 \pm 0.2) K$ with PWV expressed in mm. Not
surprisingly, the two M coefficients are different, since our
bandpasses is strongly suppressed towards the 118 GHz $O_2$ line,
while their high frequency wings pick up with small (but not
negligible) and different efficiency a residual signal from the
183 GHz $H_2O$ line. The intercept Q is sensitive to the dry-air
component, mainly to $O_2$, whose tail is observed with the same
efficiency by our detectors.

In order to account for polarized emission one has to consider
magnetic field direction and intensity. We have estimated these
for the Austral summer 2009/2010, at Dome C, by means of the
International Geomagnetic Reference Field (IGRF) model was used.
The day-by-day variation in atmospheric conditions produces a
small uncertainty on the oxygen polarized and unpolarized
brightness temperature computed through their model (seasonal
variations and magnetic storms being definitely more relevant). In
particular, in the specific case of the BRAIN campaign, the
absolute uncertainty on $O_2$ modeled signals is of the order of a
few $\mu$K and a few mK, respectively, for the V and I maps. Both
these uncertainties are negligible with respect to the one
associated with $H_2O$ emission estimates, which is of the order
of some tens of mK. This last one is driven mainly by the
day-by-day scatter of atmospheric parameters.

We found the uncertainty derived from the fits negligible with
respect to the estimated uncertainty derived from the calibration
procedure. This can be as high as 23\% and should be treated as a
systematic error. It dominates over all other uncertainties.
Calibration uncertainties proportionally propagate to the opacity
determination, while we produced monte-carlo simulations in order
to estimate PWV uncertainties.

\begin{figure}
\begin{center}
\includegraphics[angle=90,width=9cm]{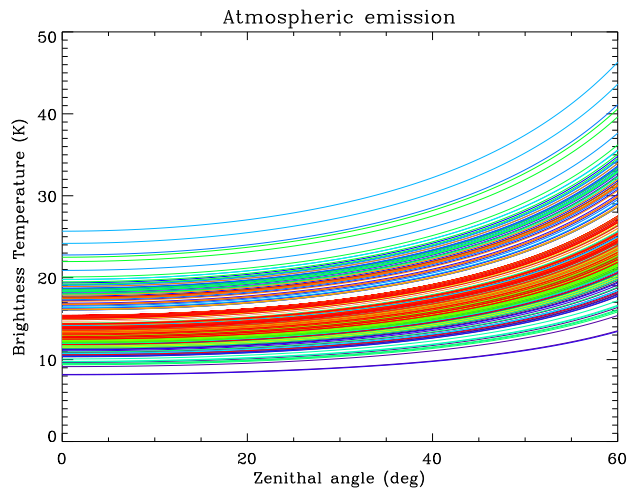}
\caption{Collection of best fit curves obtained over the
calibrated skydips collected during the 2009-2010 summer
campaign.} \label{skydip}
\end{center}
\end{figure}

In figure \ref{skydip} we report best fit curves obtained over the
calibrated skydips collected during the campaign for both
channels.

\section{Results}

\subsection{Sky opacity}

\begin{figure}
\begin{center}
\includegraphics[angle=90,width=9cm]{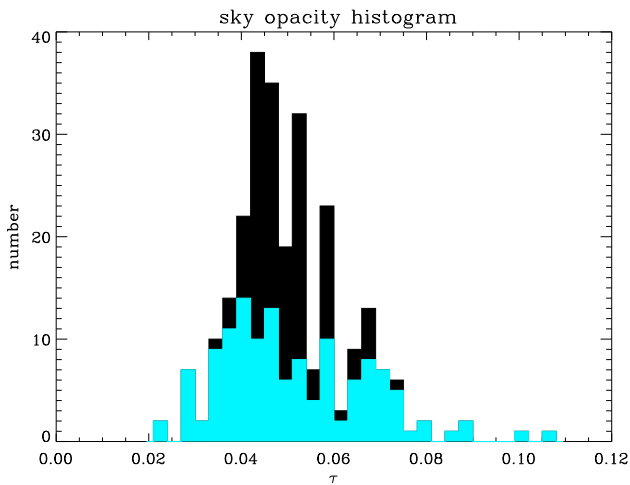}
\caption{Histogram of measured sky opacity. The cyan (light)
histogram is obtained from channel 2 measurements; the black
(dark) histogram is obtained from the collection of channel 1
measurements and is positioned over channel 2 data.}
\label{tau_histo}
\end{center}
\end{figure}

The distribution of the sky opacities measured during the campaign
can be seen in the histogram in figure \ref{tau_histo}. The cyan
(light) histogram is derived from channel 2 measurements while the
black (dark) histogram (summed and positioned on top of the cyan
one) is derived from channel 1 measurements. The measurements from
both channels result in $\tau_{I}$ values centered around an
average value $<\tau_{I}>=0.050$, with a median of
$\tau_{I}^{m}=0.048$ and statistical error on each measurement of
7\%. This corresponds to an average transmission of $95\%$.
Although all the measurements have been taken with clear sky, we
should stress that the reported results reflect a wide variety of
weather conditions and only bad weather situations (i.e. covered
sky, although rare at Dome C) have been discarded from our
analysis.

\begin{figure}
\begin{center}
\includegraphics[angle=90,width=9cm]{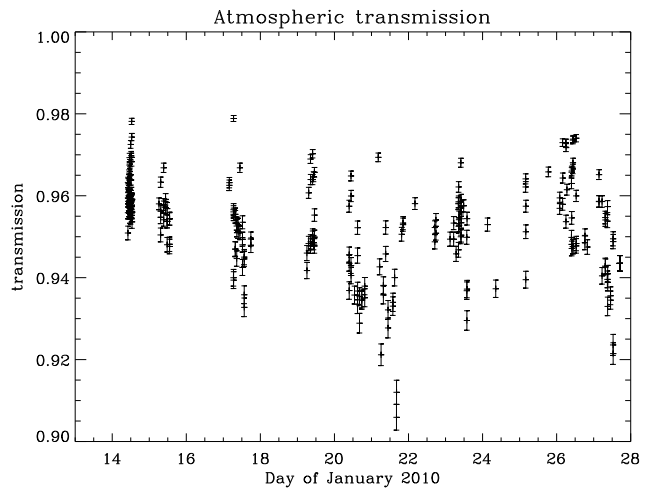}
\caption{Atmospheric transmission during the 2009-2010 campaign.
Statistical uncertainties only are shown in this graph.}
\label{trasm}
\end{center}
\end{figure}

\begin{figure}
\begin{center}
\includegraphics[angle=90,width=9cm]{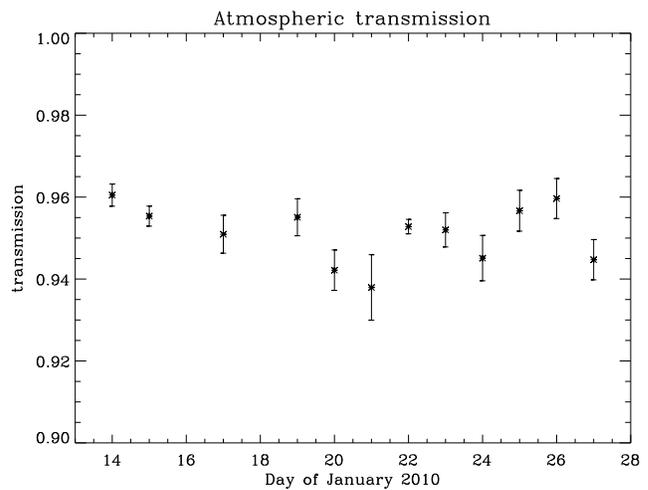}
\caption{Atmospheric transmission during the 2009-2010 campaign.
In this plot we averaged over the skydip acquired during the same
day. This shows the general trend of the sky transmission during
the campaign.} \label{trasm_campaign}
\end{center}
\end{figure}

\begin{figure}
\begin{center}
\includegraphics[angle=90,width=9cm]{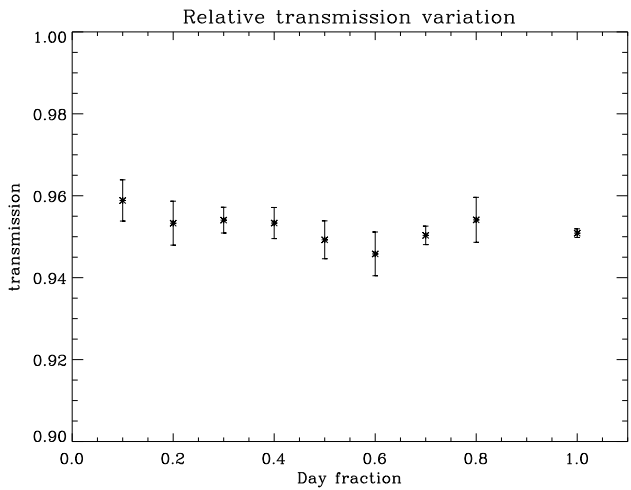}
\caption{Atmospheric transmission during the 2009-2010 campaign.
In this plot we averaged over the skydip acquired less than 2
hours apart, each measurement being normalized by the average
daily transmission, relative to (multiplied by) the overall
average transmission on the campaign. This shows the general trend
of the sky transmission during the day.} \label{trasm_day}
\end{center}
\end{figure}

In figure \ref{trasm} we show the atmospheric transmission
measurements along the whole campaign. In figure
\ref{trasm_campaign} we show the same measurements averaged on day
by day basis. Transmission seems to be fairly constant along the
campaign, with a possible general trend to decrease in the middle
of the campaign and a rise back at the end of it. Error bars in
figure \ref{trasm_campaign} reflect the variability within each
day. This parameter has to be taken into account when considering
measurements with thermal detectors like transition edge sensor
(TES) bolometers (as those planned for the QUBIC experiment) when
accounting for load variation on the bolometers themselves, and
TES plus superconducting quantum interference devices (SQUIDs)
working point tuning. We estimate a relative average variation, on
a daily basis, of 0.9 per cent, consistent with one single tuning
procedure per day needed \cite{battistelli2008}. Peaks of a few
per cent have also been observed in limited cases. This would
require tuning procedures to be run more than once a day. In
figure \ref{trasm_day} we show the transmission measured within
the day. We have averaged over skydips acquired less than 2 hours
apart on different days. In order to reduce the scatter caused by
different days of observations, we have normalized each
measurement to the average daily measurements and multiplied by
the overall average transmission. Also in this case the trend
seems to be fairly constant, with a slight decrease in the middle
of the day due to the increase of elevation of the Sun.

\subsection{Precipitable water vapor content}

We have used {\em am} model \cite{paine2011} to infer, from our
sky opacity measurements, the characteristics in terms of PWV of
the atmosphere over Dome C during our measurements. This model as
been tailored and configured for Dome C conditions and allows us
to exploit balloon data taken during the campaign, in view of the
interpretation of BRAIN data. We have found a scaling law between
a photometric quantity (the brightness temperature) and an
atmospheric parameter (PWV), thus removing possible degeneracies
in our analysis.

\begin{figure}
\begin{center}
\includegraphics[angle=90,width=9cm]{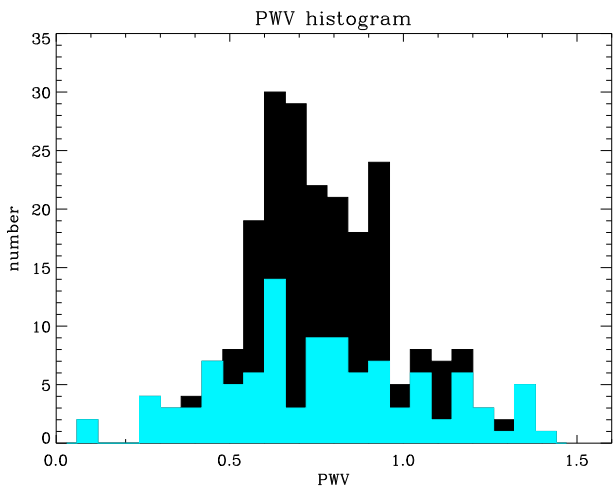}
\caption{Histogram of the derived Precipitable Water Vapor Content
on the atmosphere. The cyan (light) histogram is obtained from
channel 2 measurements; the black (dark) histogram is obtained
from the collection of channel 1 and channel 2 measurements.}
\label{PWV}
\end{center}
\end{figure}

In figure \ref{PWV} we show the histogram of the Precipitable
Water Vapor content obtained from our skydips. These are obtained
by fitting our calibrated skydips over the simulated template and
by only accounting for the brightness temperature change over the
skydip, and not its offset, and checking the zenith brightness
temperature for consistency.

During the BRAIN-pathfinder 2009-2010 campaign we find an average
PWV of $<PWV>=0.77\ mm$, with statistical errors on each
measurement of $\pm 0.06\ mm$ and calibration uncertainties of
$^{+0.15}_{-0.12}\ mm$. We find a median of $PWV^{m}=0.75\ mm$, an
average 25th percentile of $PWV^{25th}=0.49\ mm$ and an average
75th percentile of $PWV^{75th}=1.1\ mm$. This level of PWV content
is in good agreement with those measured in 2003-2005 and
2005-2009 with radio-sounding measurements \cite{tomasi2006};
\cite{tomasi2011} and with those measured by \cite{tremblin2011}
at 1.5 THz. Nevertheless, we should stress that the main results
of our measurements are obtained by directly sampling the same
spectral bandwidth of scientific interest and are thus free from
model dependent bias.

\subsection{Polarization}\label{sec:pol}

As previously mentioned, channel 2 of the BRAIN-pathfinder
measures sky emission through a rotating QWP followed by a
wire-grid polarizer. During observations we rotated the QWP at
several different speeds. For instrumental, environmental, and
noise reasons we set its physical rotational frequency at
$\nu_{QWP}=1.56\ Hz$ ($2.09\ Hz$) for most of our observations. We
thus expect any incoming circularly polarized signal to be
modulated at $ \nu_{C} = 2 \cdot \nu_{QWP} = 3.13\ Hz$ ($4.17\
Hz$) and any incoming linearly polarized signal at $ \nu_{L} = 4
\cdot \nu_{QWP} = 6.26\ Hz$ ($8.35\ Hz$) (see eq. \ref{eq:1}). The
emission of the (ambient temperature) QWP and of the polarizer
have been studied using the models presented by
\cite{salatino2011}. Our data are affected by an offset modulated
both at $\nu_{C}$ and at $\nu_{L}$ that we found to be consistent
with the QWP emission as well as the polarizer emission reflected
back from the QWP. We should stress that the stability of this
emitted signal is critical to retrieve meaningful information from
the data. One of the goals of this paper is to provide information
on the polarization of the skydip signal. For the present
analysis, it is critical to characterize and monitor the stability
of our instrument within the average time of a skydip as faster
instability will affect the results.

We have demodulated the $\nu_{C}$ and the $\nu_{L}$ signals using
band-pass filters for each raw skydip. The extracted signals have
thus been treated in the same way as the intensity signal in order
to find a possible secant law dependence in the polarized signal
of the skydips. Once we have performed secant law fits over the
polarized skydips, we can define $\tau_{C}$ and $\tau_{L}$
similarly to $\tau_{I}$:
\begin{equation}\label{eq:tau}
\centering \tau_{C} = -(B^{C}) / (C \cdot T_{0}),
\end{equation}
\begin{equation}\label{eq:tau}
\centering \tau_{L} = -(B^{L}) / (C \cdot T_{0}).
\end{equation}
The comparison between the different $\tau$'s enables us to
extract polarized information from our skydips and thus an
estimation of the air-mass-correlated (or anticorrelated)
polarization of atmospheric emission.

\begin{figure}
\begin{center}
\includegraphics[angle=90,width=9cm]{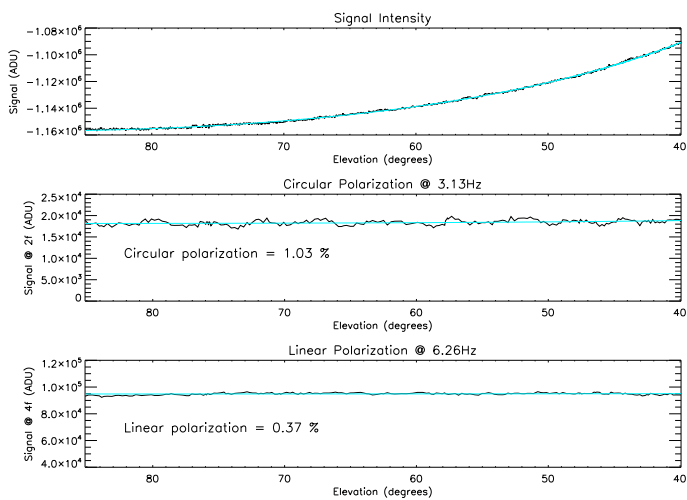}
\caption{Intensity, linear polarization and circular polarization
signal acquired during a single skydip. We plot the acquired data
(dark) and the secant law performed fit (light/cyan). We report
the derivation of the percentage polarization that can be
extracted from this skydip.} \label{pol}
\end{center}
\end{figure}

In figure \ref{pol} we show one of the skydips acquired and
analyzed in polarization. Polarization calibration has been
performed using local polarized sources placed in the near-field
and in the medium-field. We should stress, however, that the
derived percentage polarization levels are independent of the
calibration and of the determination of the equivalent temperature
of the atmosphere. The uncertainty on each of the skydip
polarization levels is thus directly derived from the fits. After
combining and weight-averaging over all the skydips, we find that
both circular and linear polarization of the air-mass correlated
signals are consistent with zero, with upper limits such that $
S_{C} < 0.19 \% $ and $ S_{L} < 0.11 \% $ (95\%CL), respectively.
These results have been confirmed by cross-correlating the
measured data with the signals expected from Zeeman splitting and
simulated using the model developed in \cite{spinelli2011}. These
limits include instrumental systematics and place a tight limit on
the QWP plus polarizer stability within the time of the acquired
skydips.

\section{Conclusions}

In this paper we report a detailed site testing of Dome C at $150\
GHz$ and the first limits on the polarized emission of its
atmosphere. Dome Concordia is demonstrated to be an exceptional
millimetric observational site, in terms of absolute transmission
and stability and polarization limits are encouraging as far as
spurious polarization and intensity-polarization mixing are
concerned. Our opacity measurements have been derived from direct
sampling of the frequencies of astronomical interest and are thus
free from model-dependent bias. We anticipate that, in most
observational conditions, the measured daily stability should
enable us to have Transition Edge Sensor Bolometer arrays
requiring only a single tuning procedure per day. When comparing
our millimetric opacities and in-bandwidth transmissions with
those obtained at sub-millimeter wavelengths, the absolute values
are one order of magnitude better in terms of transmission and
stability. Nevertheless, we should keep in mind that the
requirements, in terms of systematic control and stability, for a
millimetric B-modes CMB experiment, are such that it is necessary
to characterize the atmospheric stability to a high level of
precision and our results are very encouraging. Our derived PWV
value relies on a model independently developed by the BRAIN
collaboration and tailored for Dome C. Our PWV values are
consistent with those derived at sub-millimeter wavelenghts
\cite{tomasi2011}. In our case, however, radiation is detected
through an instrument similar to those aiming to detect CMB
polarization, allowing direct monitoring of many systematic
effects. Systematic control requirements for a B-mode CMB
experiment are so stringent that no instrument, to date, has been
able to meet them. The requirement for the instrumental spurious
polarization induced from leakage of CMB anisotropies into B-mode
polarization is that the leakage should be maintained lower than
$10^{-3}$ \cite{bock2006}. This is necessary in order to be able
to reach a B-modes signal of the order of $30\ nK$ rms (i.e.
Tensor to Scalar ratio r=0.01). In terms of absolute temperature
limits, our analysis does not allow us to reach this limit.
However, in terms of a relative systematic limit, considering that
our 0.1\% limit for linear polarization includes spurious
polarization due to intensity-to-polarization leakage, our
instrument is already satisfying to this requirement.

\section*{Acknowledgments}
This work is supported and funded by the ``Progetto Nazionale
Ricerche in Antartide'' (PNRA) and the ``Institut Polaire
francaise Paul Emile Victor'' (IPEV). We thank the logistic
support at Dome C. We acknowledge Dr. Andrea Pellegrini for the
radio sound data and information obtained from IPEV/PNRA Project
``Routine Meteorological Observation at Station Concordia -
www.climantartide.it.'' We thank Ken Ganga for comments and for
reviewing the paper. We acknowledge Scott Paine (Smithsonian
Astrophysical Observatory) for making the {\em am} code available
and for the kind support. We acknowledge the anonymous referee for
comments that improved the paper.

\label{lastpage}

\end{document}